# Size-dependent couple stress Timoshenko beam theory

Ali R. Hadjesfandiari, Arezoo Hajesfandiari, Haoyu Zhang, Gary F. Dargush

Department of Mechanical and Aerospace Engineering
University at Buffalo, The State University of New York, Buffalo, NY 14260 USA

ah@buffalo.edu, ah62@buffalo.edu, haoyuzha@buffalo.edu, gdargush@buffalo.edu

December 20, 2017

**Abstract**

In this paper, a new size-dependent Timoshenko beam model is developed based on the consistent couple stress theory. In the present formulation, the governing equations and corresponding boundary conditions are obtained. Afterwards, this formulation is used to investigate size-dependency for several elementary beam problems. Analytical solutions are obtained for pure bending of a beam and for a cantilever beam with partially and fully clamped boundary conditions. These analytical results are then compared to the numerical results from a two-dimensional finite element formulation for the corresponding couple stress continuum problem.

**1. Introduction**

It is well known that classical continuum mechanics cannot predict the behavior of materials for very small length scales. While molecular mechanics theories have certainly enjoyed some success, these approaches are only computationally feasible for collections of particles of quite limited spatial and temporal extent. This is the true motivation for developing a size-dependent continuum theory. Mindlin and Tiersten [1] and Koiter [2] developed an initial incomplete version of couple stress theory (MTK-CST), which suffers from some fundamental inconsistencies, such as the indeterminacy of the couple-stress tensor, the appearance of the normal component of couple-traction vector on boundary surfaces, and the consideration of the redundant body couple distribution [3-7]. For linear isotropic elastic materials this theory requires two couple-stress material parameters, which does not seem attractive from a practical point of view. It turns out that only one of these elastic couple-stress coefficients appears in the final governing equations when written in terms of displacements. This in hindsight demonstrates the inconsistency of MTK-CST.

Without resolving the inconsistencies of MTK-CST, Yang et al. [8] violated fundamental rules of mechanics to reduce the number of couple-stress material parameters in this theory for linear isotropic elastic material from two coefficients to only one. In their development, they introduced an extra artificial equilibrium equation for the moment of couples that has no physical reality, but apparently makes the couple-stress tensor symmetric. It turns out this proposed theory, called modified couple stress theory (M-CST), suffers from the same fundamental inconsistencies as MTK-CST. Furthermore, in this still indeterminate theory (M-CST), the symmetric couple-stress tensor has a torsional character, which results in size effects for torsional and anticlastic deformation, but not for bending.

From a practical point of view, even if we ignore the indeterminacy of the couple-stress tensor, it is generally impossible to satisfy all boundary conditions correctly in many problems using the original MTK-CST and M-CST. Consequently, MTK-CST and M-CST are not suitable theories within continuum mechanics for developing new size-dependent formulations. These theories predict certain deformations, which contradict common sense and do not agree with experiments. This can be observed in very elementary practical problems. For example, there is no consistent solution for pure torsion of a circular bar in these theories. We notice that the inconsistent approximate solutions for pure torsion in these theories predict significant size effect, which does not agree with experiments [9]. MTK-CST and M-CST also cannot describe pure bending of a plate properly [10]. Particularly, M-CST predicts no couple-stresses and no size effect for the pure bending of the plate into a spherical shell. Consequently, MTK-CST and M-CST should not be used anymore to describe physical reality.

Hadjesfandiari and Dargush [3] and Hadjesfandiari et al. [4] have resolved all inconsistencies and confusions in the original couple stress theory (MTK-CST) and developed the consistent couple-stress theory (C-CST). In this theory, the couple-stress tensor is skew-symmetric and is energetically conjugate to the skew-symmetric mean curvature tensor or mean curvature rate tensor for solids and fluids, respectively. It turns out that the skew-symmetric couple-stress tensor has a vectorial character, and results in a size effect for bending deformation. For the linear isotropic elastic solid, there is only one additional material property, $l$, with the dimensions of length, which becomes important for problems having characteristic geometry on the order of $l$ or smaller.

It should be emphasized that C-CST is not a special case of the original MTK-CST in a physical sense, because a consistent theory should never be classified as a special case of an inconsistent theory. As we have mentioned, MTK-CST and M-CST suffer from many mathematical and physical inconsistencies. We also notice that C-CST uses a different curvature tensor from the original MTK-CST. Contrary to MTK-CST and M-CST, the consistent couple stress theory (C-CST) predicts consistent results for pure torsion of a circular bar [9] and pure bending of a plate [10]. Over the last several decades, many different continuum theories have been proposed. However, only C-CST satisfies all criteria necessary for a consistent size-dependent continuum mechanics. Therefore, it provides a powerful tool to develop new formulations for different coupled multi-physics problems, such as piezoelectricity [11] and thermoelasticity [12]. This theory has also been introduced into fluid mechanics to model size-dependency and perhaps to contribute to the understanding of turbulence, which affects a cascade of length scales [4].

Consistent couple stress theory (C-CST) has been used recently to study the size effect in some elastodynamical problems, see Ref. [13-17]. Salter and Richardson [18] developed the governing equations for the equilibrium of smoothly heterogeneous couple-stress materials by using the extended Hamilton's principle. The application of this new consistent couple stress theory in fluid dynamics can also be found in Ref. [19-21]. However, the number of analytical solutions available for C-CST within the context of elasticity and viscous fluid is very limited, and therefore, approximate techniques must be explored. Computational methods such as finite element and boundary element methods for solving linear two-dimensional couple stress problems have already been developed, see Ref. [22-30]. Furthermore, the finite difference method in the framework of size-dependent fluid mechanics has also been implemented by Hajesfandiari et al. [31]. However, for three-dimensional cases, the formulations become formidable and, consequently, other computational methods should be developed.

Structural mechanics methods offer another possibility to analyze size effects in micro- and nano-beams, plates and shells. The self-consistency of C-CST makes it suitable for developing size-dependent structural models, such as beams, plates and shells. However, it should be noticed that the size-dependent modeling based on structural mechanics methods requires more approximation than the classical structural modeling. As a result, this approach should be used with more caution.

There have been some recent structural formulations based on C-CST. Alashti and Abolghasemi [32] have developed an Euler-Bernoulli model to analyze static and free vibration of micro-beams. Fakhrabadi [33,34] and Fakhrabadi and Yang [35] have investigated the static and dynamic electromechanical behavior of carbon nano-tubes and nano-beams by using linear and a non-linear Euler-Bernoulli beam model. Li et al. [36] have used a three-layer Euler-Bernoulli micro-beam model to study the size-dependent flexoelectric effect under static and dynamic conditions. Beni [37,38] has studied static deformation, buckling and free vibration of piezoelectric nano-beams by using linear and non-linear C-CST size-dependent Timoshenko beam models. Keivani et al. [39-41] have used C-CST based Euler-Bernoulli beam model to investigate the dynamic stability of beam-type nanotweezers, paddle-type and double-sided NEMS measurement sensors. Zozulya [42] has developed size-dependent Timosheko and Euler-Bernoulli models for curved rods based on C-CST. Nejad et al. [43] have investigated free vibration nano-beams made of arbitrary bi-directional functionally graded materials by using a C-CST Euler-Bernoulli beam mode. Ji and Li [44] have developed a size-dependent flexoelectric model of Kirchhoff-Love plate bending based on C-CST to study circular micro-plates in static and dynamic conditions. Aghababaie Beni et al. [45] have used C-CST to develop a size-dependent plate model to study the dynamic response of microplates. Additionally, Kheibari and Beni [46], Razavi et al. [47] and Dehkordi and Beni [48] have analyzed free vibration of single-walled piezoelectric nanotubes, functionally graded piezoelectric cylindrical nano-shell and nano-cone by using C-CST.

It should be mentioned that there also are many structural formulations based on the other size-dependent theories, such as modified couple stress theory (M-CST) [8]. For example, Park and Gao [49], Kong et al. [50], Ma et al. [51], Asghari et al. [52,53], Reddy [54], Li et al. [55], Chen and Meguid [56], Gao [57], Karttunen et al. [58] and Goncalves et al. [59] have developed different size-dependent beam models. It turns out that the in-plane solutions for M-CST are similar to those in C-CST and can be found by scaling $l \to l/2$. However, the apparent success of M-CST in describing size effect for isotropic elastic beam bending is not enough to justify M-CST as a correct theory. It should be noticed that for beam bending the in-plane solutions from all couple stress theories (MTK-CST, M-CST and C-CST) are the same, but out-of-plane solutions are different. Beam formulations in these theories are approximate structural analysis methods, which are expressed by ordinary differential equations for the static case. As a result, these beam models cannot demonstrate the validity of any of these theories for general three-dimensional boundary value problems. As previously mentioned, the couple-stresses in M-CST create anticlastic deformation, which is equivalent to a torsion, not bending. As a result, modified couple stress theory (M-CST) cannot describe the bending of plates properly [10]. In beam and plate theories based on M-CST, the normal force-stresses create bending deformation, whereas the couple-stresses create torsional deformation. Therefore, the apparent success of M-CST for beam formulations has been very misleading. It is regrettable to see that users of this theory do not recognize these serious inconsistencies.

In the present work, we utilize consistent couple stress theory (C-CST) and develop a size-dependent Timoshenko beam model. In this formulation, the governing equations and corresponding boundary conditions are derived. Then, the formulation is used to investigate the size-dependent effect for several specific beam problems. Analytical solutions for pure bending of a beam and for a cantilever beam with partially and fully clamped boundary conditions are obtained. A number of limiting forms are also explored, including the remarkable inflexurable case in which bending deformation is entirely suppressed. Many of these analytical results are then compared to the numerical results from a two-dimensional finite element formulation.

The balance of this paper is structured as follows. In Section 2, we provide an overview of consistent couple stress theory (C-CST). In Section 3, we develop the size-dependent Timoshenko beam model based on C-CST, including details on the governing equations and corresponding boundary conditions. In Section 4, this new model is used to investigate size-dependency in pure bending of a beam, while the bending of a cantilever beam is studied in Section 5. The latter case includes two distinct partially and fully clamped sets of boundary conditions. Afterwards, in Section 6, we compare the analytical results to the numerical results from a two-dimensional finite element formulation for the underlying continuum theory. Finally, we offer some conclusions in Section 7.

## 2. Consistent couple stress theory

In couple stress theory, the interaction in the body is represented by force-stress $\sigma_{ij}$ and couple-stress $\mu_{ij}$ tensors. The force and moment balance equations for general couple stress theory under quasi-static conditions in the absence of body forces are written, respectively, as:

$$\sigma_{ji,j} = 0 \tag{1}$$

$$\mu_{ji,j} + \varepsilon_{ijk}\sigma_{jk} = 0 \tag{2}$$

where $\varepsilon_{ijk}$ is the Levi-Civita alternating symbol.

The force-stress tensor $\sigma_{ij}$ is generally non-symmetric and can be decomposed as

$$\sigma_{ij} = \sigma_{(ij)} + \sigma_{[ij]} \tag{3}$$

where $\sigma_{(ij)}$ and $\sigma_{[ij]}$ are the symmetric and skew-symmetric parts, respectively. Based on the consistent couple stress theory (C-CST) [3], the couple-stress tensor $\mu_{ij}$ is skew-symmetrical

$$\mu_{ji} = -\mu_{ij} \tag{4}$$

The true couple-stress vector $\mu_i$ dual to the pseudo-tensor $\mu_{ij}$ is defined as

$$\mu_i = \frac{1}{2}\varepsilon_{ijk}\mu_{kj} \tag{5}$$

For the kinematics, the infinitesimal strain and rotation tensors are defined as

$$e_{ij} = \frac{1}{2}\left(u_{i,j} + u_{j,i}\right) \tag{6}$$

$$\omega_{ij} = \frac{1}{2}\left(u_{i,j} - u_{j,i}\right) \tag{7}$$

respectively. Since the true tensor $\omega_{ij}$ is skew-symmetrical, one can introduce its corresponding dual pseudo rotation vector as

$$\omega_i = \frac{1}{2}\varepsilon_{ijk}\omega_{kj} \tag{8}$$

The infinitesimal skew-symmetric mean curvature tensor is defined as

$$\kappa_{ij} = \omega_{[i,j]} = \frac{1}{2}\left(\omega_{i,j} - \omega_{j,i}\right) \tag{9}$$

For the most general linear anisotropic elastic material, the constitutive relations are [3]

$$\sigma_{(ji)} = A_{ijkl}e_{kl} + C_{ijkl}\kappa_{kl} \tag{10}$$

$$\mu_{ji} = B_{ijkl}\kappa_{kl} + C_{klij}e_{kl} \tag{11}$$

For a linear isotropic elastic material, these constitutive tensors reduce to

$$A_{ijkl} = \lambda\delta_{ij}\delta_{kl} + G\delta_{ik}\delta_{jl} + G\delta_{il}\delta_{jk} \tag{12}$$

$$B_{ijkl} = 4Gl^2\left(\delta_{ik}\delta_{jl} - \delta_{il}\delta_{jk}\right) \tag{13}$$

$$C_{ijkl} = 0 \tag{14}$$

Therefore, in the linear isotropic size-dependent consistent couple stress elasticity, the constitutive relations are

$$\sigma_{(ij)} = \lambda e_{kk}\delta_{ij} + 2Ge_{ij} \tag{15}$$

$$\mu_{ij} = -8Gl^2\kappa_{ij} \tag{16}$$

Here the moduli $\lambda$ and $G$ are the Lamé constants for isotropic media in Cauchy elasticity, and $G$ is also referred to as the shear modulus. These two constants are related by

$$\lambda = 2G\frac{\nu}{1-2\nu} \tag{17}$$

where $\nu$ is the Poisson ratio. In addition, we have the relations

$$\lambda = \frac{\nu E}{(1+\nu)(1-2\nu)} \qquad G = \frac{E}{2(1+\nu)} \qquad 3\lambda + 2G = \frac{E}{1-2\nu} \qquad (18\text{-}20)$$

where $E$ is Young's modulus of elasticity. The constant $l$ is the characteristic material length scale parameter in the consistent couple stress theory (C-CST), which is absent in Cauchy elasticity, but is fundamental to small deformation couple stress elasticity.

Furthermore, from (2), we derive

$$\sigma_{[ji]} = -\mu_{[i,j]} = 2Gl^2 \nabla^2 \omega_{ji} \qquad (21)$$

As a result, for the total force-stress tensor, we have

$$\sigma_{ji} = \lambda e_{kk} \delta_{ij} + 2G e_{ij} + 2Gl^2 \nabla^2 \omega_{ji} \qquad (22)$$

while the elastic energy density function $W$ for material is given by

$$W = \frac{1}{2}\lambda (e_{kk})^2 + G e_{ij} e_{ij} + 4Gl^2 \kappa_{ij} \kappa_{ij} \qquad (23)$$

## 3. Size-dependent Timoshenko beam with couple-stresses

In this section, the couple stress theory is used to reformulate the Timoshenko beam problem to account for size-dependency in a beam bent in one of its principal planes. For simplicity we may assume the symmetrical bending in the vertical symmetry plane $xz$ of the beam. The governing equilibrium equations for an infinitesimal element of the beam, as shown in Fig. 1, are

$$\frac{dQ}{dx} + q = 0 \qquad (24)$$

$$\frac{dM}{dx} + Q = 0 \qquad (25)$$

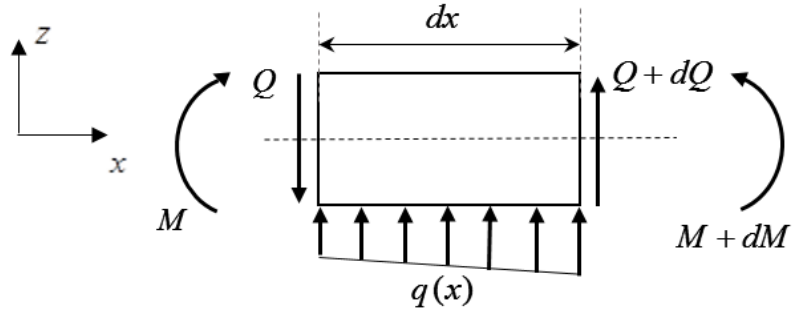

**Fig. 1.** Force-moment system for an infinitesimal element of beam.

where $Q$ and $M$ are total transverse force and bending moment, and $q$ is the distributed transverse load.

In Timoshenko beam theory, the displacement field is assumed to be

$$u = -z\beta(x) \tag{26}$$

$$v = 0 \tag{27}$$

$$w = w(x) \tag{28}$$

where $\beta$ denotes the rotation of the cross-section at the mid-plane, and $w$ is the transverse deflection of the mid-plane of the beam, as shown in Fig. 2. Here $\theta$ is the slope of the central axis of the beam, where

$$\frac{dw}{dx} = \theta \tag{29}$$

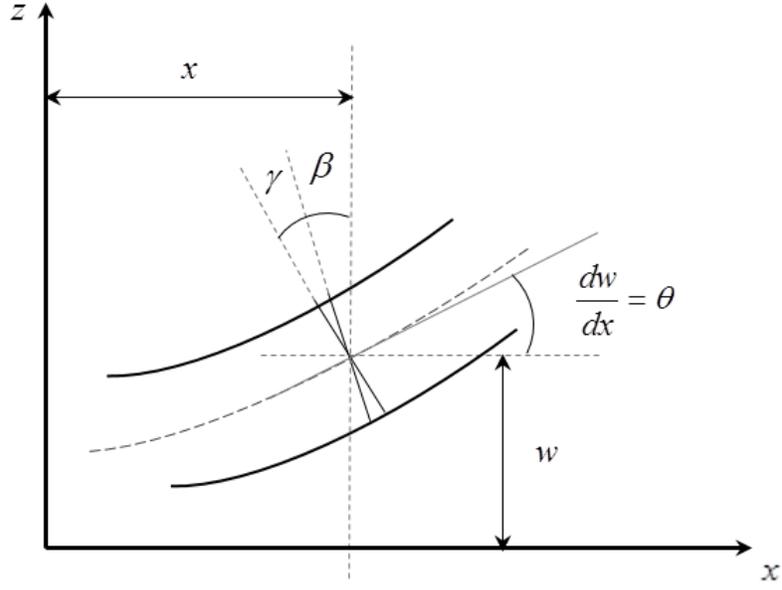

**Fig. 2.** Deformation and coordinate system.

The non-zero strains, rotation and curvature in this theory are, respectively,

$$e_{xx} = \frac{du}{dx} = -z\frac{d\beta}{dx} \tag{30}$$

$$e_{zx} = \frac{1}{2}\left(\frac{dw}{dx} - \beta\right) \tag{31}$$

$$\omega_y = -\frac{1}{2}\left(\frac{dw}{dx} + \beta\right) \tag{32}$$

$$\kappa_{xy} = -\kappa_z = -\frac{1}{2}\frac{\partial \omega_y}{\partial x} = \frac{1}{4}\left(\frac{d^2w}{dx^2} + \frac{d\beta}{dx}\right) \tag{33}$$

It is convenient to define the engineering rotation $\phi$, engineering shear strain $\gamma$ and engineering curvature $\mathrm{K}$ as

$$\phi = -\omega_y = \frac{1}{2}\left(\frac{dw}{dx} + \beta\right) \tag{34}$$

$$\gamma = 2e_{zx} = \frac{dw}{dx} - \beta \tag{35}$$

$$\mathrm{K} = \frac{d\phi}{dx} = -2\kappa_z = \frac{1}{2}\left(\frac{d^2w}{dx^2} + \frac{d\beta}{dx}\right) \tag{36}$$

According to one-dimensional constitutive relations for slender beams, the non-zero stresses are

$$\sigma_{xx} = Ee_{xx} = -E\frac{d\beta}{dx}z \tag{37}$$

$$\sigma_{(xz)} = G\gamma = G\left(\frac{dw}{dx} - \beta\right) \tag{38}$$

$$\mu_z = -\mu_{xy} = 4Gl^2\mathrm{K} = 2Gl^2\left(\frac{d^2w}{dx^2} + \frac{d\beta}{dx}\right) \tag{39}$$

Therefore

$$\sigma_{[xz]} = -\frac{1}{2}\frac{d\mu_z}{dx} = -2Gl^2\frac{d\mathrm{K}}{dx} = -Gl^2\frac{d}{dx}\left(\frac{d^2w}{dx^2} + \frac{d\beta}{dx}\right) \tag{40}$$

However, we should notice that the transverse force-stress and couple-stress components are not uniformly distributed on the cross-section of the beam.

The bending moment $M$ and transverse force $Q$ on the cross-section of the beam can be decomposed as

$$M = M_\sigma + M_\mu \tag{41}$$

$$Q = Q_\sigma + Q_\mu \tag{42}$$

where

$$M_\sigma = -\int_A z\sigma_{xx}dA \tag{43}$$

$$M_\mu = -\int_A \mu_{xy}dA \tag{44}$$

and

$$Q_\sigma = \int_A \sigma_{(xz)} dA \tag{45}$$

$$Q_\mu = \int_A \sigma_{[xz]} dA \tag{46}$$

It should be emphasized that since the couple-stress tensor is skew-symmetric, $\mu_{yx} = -\mu_{xy}$, the couple-stress moment $M_\mu$ creates bending deformation. However, in the modified couple stress theory (M-CST), where the couple-stress tensor is symmetric, $\mu_{yx} = \mu_{xy}$, the couple-stress moment creates torsional deformation [10]. This clearly demonstrates a major inconsistency of M-CST.

We notice that

$$Q_\mu = \int_A \sigma_{[xz]} dA = -\int_A \frac{1}{2} \frac{\partial \mu_z}{\partial x} dA = -\frac{1}{2} \frac{d}{dx} \int_A \mu_z dA \tag{47}$$

which shows

$$Q_\mu = -\frac{1}{2} \frac{dM_\mu}{dx} \tag{48}$$

By using the constitutive relations, we obtain

$$M_\sigma = EI \frac{d\beta}{dx} \tag{49}$$

$$M_\mu = 4k_\mu GAl^2 \mathrm{K} = 2k_\mu GAl^2 \left( \frac{d^2w}{dx^2} + \frac{d\beta}{dx} \right) \tag{50}$$

$$Q_\sigma = k_s GA\gamma = k_s GA \left( \frac{dw}{dx} - \beta \right) \tag{51}$$

and as a result

$$Q_\mu = -\frac{1}{2} \frac{dM_\mu}{dx} = -\frac{d}{dx} \left[ k_\mu GAl^2 \left( \frac{d^2w}{dx^2} + \frac{d\beta}{dx} \right) \right] \tag{52}$$

We notice that components $M_\sigma$ and $Q_\sigma$ are the bending moment and transverse force due to the normal force-stress and symmetric shear force-stress, respectively, as in classical Timoshenko beam theory. However, there are additional components $M_\mu$ and $Q_\mu$, which are due to the effect of couple-stress $\mu_{xy}$ and skew-symmetric shear force-stress $\sigma_{[xz]}$, respectively, in the couple-stress Timoshenko beam theory. The transverse component force $Q_\mu$ can be called the couple-stress induced transverse force. Here $A$ and $I$ are the area and second moment of area of the beam cross-section, and the coefficient $k_s$ is the shear coefficient, which accounts for the non-uniformity of the symmetric shear stress $\sigma_{(xz)}$ over the beam cross-section. We have also introduced the correction factor $k_\mu$ to account for the non-uniformity of the couple stress $\mu_{xy}$ over the beam cross-section. It should be noticed that in couple stress theory the coefficients $k_s$ and $k_\mu$ depend on Poisson ratio $\nu$ and the length scale parameter $l$. However, we still use the value of $k_s$ from the classical theory [60] as an approximation.

It should be noticed that the results in this formulation can be used for the cylindrical bending of a wide flat plate with width $b$ and height $h$, when $b \gg h$. It is only necessary to replace the Young's modulus $E$ by $\dfrac{E}{1-\nu^2}$ in (49) and take $k_\mu = 1$.

Consequently, we obtain the following relations for the total bending moment $M$ and the transverse force $Q$:

$$M = EI \frac{d\beta}{dx} + 2k_\mu G A l^2 \left( \frac{d^2 w}{dx^2} + \frac{d\beta}{dx} \right) \tag{53}$$

$$Q = k_s G A \left( \frac{dw}{dx} - \beta \right) - \frac{d}{dx}\left[ k_\mu G A l^2 \left( \frac{d^2 w}{dx^2} + \frac{d\beta}{dx} \right) \right] \tag{54}$$

By substituting the expressions for $Q$ and $M$ in the equilibrium equations (24) and (25), we obtain the equilibrium equations

$$\frac{d}{dx}\left\{ k_s G A \left( \frac{dw}{dx} - \beta \right) - \frac{d}{dx}\left[ k_\mu G A l^2 \left( \frac{d^2 w}{dx^2} + \frac{d\beta}{dx} \right) \right] \right\} + q = 0 \tag{55}$$

$$\frac{d}{dx}\left[EI\frac{d\beta}{dx}+k_\mu GAl^2\left(\frac{d^2w}{dx^2}+\frac{d\beta}{dx}\right)\right]+k_s GA\left(\frac{dw}{dx}-\beta\right)=0 \tag{56}$$

The total potential energy for this Timoshenko beam with couple-stresses is given by

$$\Pi=\int_0^L U dx-\int_0^L qwdx-[M_\sigma\beta]_0^L-[M_\mu\phi]_0^L-[Qw]_0^L \tag{57}$$

where

$$U=\frac{1}{2}EI\left(\frac{d\beta}{dx}\right)^2+\frac{1}{2}k_s GA\gamma^2+2k_\mu GAl^2\mathrm{K}^2 \tag{58}$$

is the elastic energy per unit length of the deformed beam, and the term $-\int_0^L qwdx$ is the potential of the transverse load $q$. We notice that the terms $-[M_\sigma\beta]_0^L$, $-[M_\mu\phi]_0^L$ and $-[Qw]_0^L$ are the potential of the force-stress bending moment $M_\sigma$, the couple-stress bending moments $M_\mu$, and the total transverse force $Q$ at the ends of the beam, respectively. Consequently, the total potential energy can be written as

$$\Pi=\int_0^L\left[\frac{1}{2}EI\left(\frac{d\beta}{dx}\right)^2+\frac{1}{2}k_s GA\left(\frac{dw}{dx}-\beta\right)^2+\frac{1}{2}k_\mu GAl^2\left(\frac{d^2w}{dx^2}+\frac{d\beta}{dx}\right)^2\right]dx$$
$$-\int_0^L pwdx-[M_\beta\beta]_0^L-\left[M_\theta\frac{dw}{dx}\right]_0^L-[Qw]_0^L \tag{59}$$

where we have used the relation $\phi=\frac{1}{2}\left(\frac{dw}{dx}+\beta\right)$. We notice that the total bending moment on the cross-section of the beam has been decomposed to two components $M_\beta$ and $M_\theta$, where

$$M_\beta=M_\sigma+\frac{1}{2}M_\mu=EI\frac{d\beta}{dx}+k_\mu GAl^2\left(\frac{d^2w}{dx^2}+\frac{d\beta}{dx}\right) \tag{60}$$

$$M_\theta=\frac{1}{2}M_\mu=k_\mu GAl^2\left(\frac{d^2w}{dx^2}+\frac{d\beta}{dx}\right) \tag{61}$$

Therefore, the transverse displacement degree of freedom $w$ is energy conjugate to the total transverse force $Q$, where

$$Q = Q_\sigma + Q_\mu \quad \leftrightarrow \quad w \tag{62}$$

It is also seen that in this formulation, the slope $\theta = \dfrac{dw}{dx}$ and rotation $\beta$ are independent degrees of freedom, and their conjugate bending moments are

$$\begin{cases} M_\beta = M_\sigma + \dfrac{1}{2} M_\mu & \leftrightarrow \quad \beta \\ M_\theta = \dfrac{1}{2} M_\mu & \leftrightarrow \quad \theta = \dfrac{dw}{dx} \end{cases} \tag{63}$$

The equilibrium corresponds to equating the first variation to zero, that is,

$$\delta \Pi = 0 \tag{64}$$

This is the weak formulation or virtual work theorem for this model, which can be written as

$$\int_0^L EI \left( \frac{d\beta}{dx} \right) \delta \left( \frac{d\beta}{dx} \right) dx + \int_0^L GAk_s \gamma \delta \gamma \, dx + \int_0^L 4GAk_\mu l^2 \mathcal{K} \delta \mathcal{K} \, dx$$
$$= \int_0^L p \delta w \, dx + \left[ M_\beta \delta \beta \right]_0^L + \left[ M_\theta \delta \theta \right]_0^L + \left[ Q \delta w \right]_0^L \tag{65}$$

It is seen that for the essential (geometrical) boundary conditions, we can specify $w(x)$, $\theta = \dfrac{dw}{dx}$ or $\beta(x)$. For the natural boundary conditions, we may specify $Q$, $M_\theta$ or $M_\beta$.

It should be noticed that in classical Timoshenko beam theory $w(x)$ and $\beta(x)$ are the kinematical variables or the degrees of freedom, whereas the slope $\theta = \dfrac{dw}{dx}$ is not a fundamental variable. However, in size-dependent couple stress Timoshenko theory, $\theta = \dfrac{dw}{dx}$ becomes a degree of freedom, which accounts for a portion of the engineering rotation $\phi = \frac{1}{2}\left(\frac{dw}{dx}+\beta\right)$. This shows that the couple stress Timoshenko beam theory combines the features of classical Timoshenko and Euler-Bernoulli beam theories.

Now we use this formulation to obtain analytical solutions for the uniform cross-section micro-beam under different loadings conditions. We notice that for a beam with uniform cross-section, the total bending moment $M$ and the shear force $Q$ become

$$M = \left(EI + 2k_\mu GAl^2\right)\frac{d\beta}{dx} + 2k_\mu GAl^2 \frac{d^2 w}{dx^2} \tag{66}$$

$$Q = GA\left(k_s \frac{dw}{dx} - k_\mu l^2 \frac{d^3 w}{dx^3}\right) - GA\left(k_s \beta + k_\mu l^2 \frac{d^2\beta}{dx^2}\right) \tag{67}$$

## 4. Pure bending of a beam

Consider the pure bending of the beam with uniform cross-section, as shown in Fig. 3. For this loading, the distribution of bending moment $M$ and transverse force $Q$ are

$$M(x) = M_0, \quad Q(x) = 0 \tag{68a-b}$$

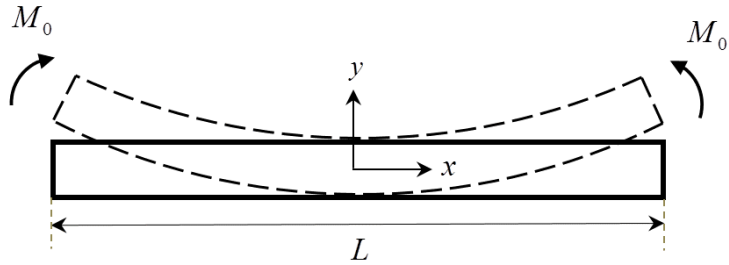

**Fig. 3.** Pure bending.

Interestingly, the deformation is exactly the same as the deformation in couple stress Euler-Bernoulli beam theory with the flexural rigidity $EI + 4k_\mu GAl^2$, where up to an arbitrary vertical rigid motion

$$w(x) = \frac{M_0}{2\left(EI + 4k_\mu GAl^2\right)} x^2 \tag{69}$$

$$\beta = \frac{dw}{dx} = \frac{M_0}{EI + 4k_\mu GAl^2} x \tag{70}$$

We notice that this deformation corresponds to the fully clamped condition at $x = 0$, where

$$w = 0, \quad \beta = 0, \quad \frac{dw}{dx} = \theta = 0 \tag{71a-c}$$

For this deformation there is no engineering shear strain, $\gamma = 0$, and the continuum mechanical rotation $\phi$ and engineering curvature $\mathcal{K}$ become

$$\phi = \frac{1}{2}\left(\frac{dw}{dx} + \beta\right) = \frac{M_0}{EI + 4k_\mu GAl^2} x \tag{72}$$

$$\mathcal{K} = \frac{d\phi}{dx} = \frac{M_0}{EI + 4k_\mu GAl^2} \tag{73}$$

For the force- and couple-stress bending moments, we obtain

$$M_\sigma = EI \frac{d\beta}{dx} = \frac{EI}{EI + 4k_\mu GAl^2} M_0 \tag{74}$$

$$M_\mu = 4k_\mu GAl^2 \mathcal{K} = \frac{4k_\mu GAl^2}{EI + 4k_\mu GAl^2} M_0 \tag{75}$$

where

$$M_\sigma + M_\mu = M_0 \tag{76}$$

However, the shear and couple-stress induced transverse forces vanish, that is, $Q_\sigma = Q_\mu = 0$.

We notice that as $l$ increases, or the beam become more slender, the couple stress bending moment $M_\mu$ increases. This shows contrary to classical beam theory, for micro-beams, the load is carried more by couple-stresses than normal force-stresses.

## 5. Cantilever beam

Now we consider the bending of the cantilever beam with uniform cross-section under the end transverse force $P$, shown in Fig. 4.

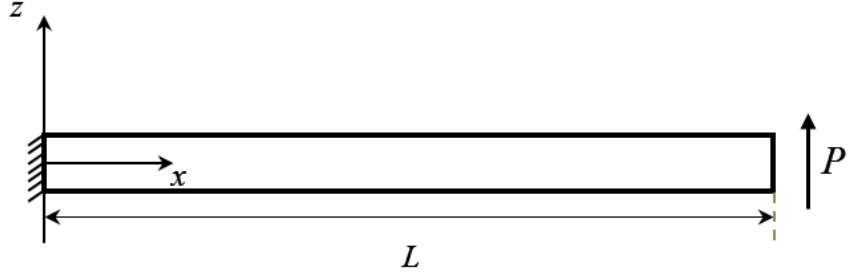

**Fig. 4.** Cantilever beam.

For this loading the distribution of transverse force $Q$ and bending moment $M$ are

$$Q = P \tag{77}$$

$$M = P(L - x) \tag{78}$$

Therefore, the relations (53) and (54) become

$$M = \left(EI + 2k_\mu GAl^2\right)\frac{d\beta}{dx} + 2k_\mu GAl^2 \frac{d^2 w}{dx^2} = P(L - x) \tag{79}$$

$$Q = GA\left(k_s \frac{dw}{dx} - k_\mu l^2 \frac{d^3 w}{dx^3}\right) - GA\left(k_s \beta + k_\mu l^2 \frac{d^2 \beta}{dx^2}\right) = P \tag{80}$$

We notice that the boundary conditions at the right end at $x = L$ are

$$Q = P, \quad M_\sigma = 0, \quad M_\mu = 0 \tag{81a-c}$$

However, we can consider two types of boundary conditions at the left end, $x = 0$, of the cantilever:

1. Partially clamped at $x = 0$, where

$$w = 0, \quad \beta = 0, \quad M_\mu = 0 \tag{82a-c}$$

2. Fully clamped at $x = 0$, where

$$w = 0, \qquad \beta = 0, \qquad \theta = \frac{dw}{dx} = 0 \qquad (83\text{a-c})$$

By integrating the relation (79) once and noticing that $\beta = 0$ and $w = 0$ at $x = 0$ in both cases, we obtain

$$\left(EI + 2k_\mu GAl^2\right)\beta + 2k_\mu GAl^2\left(\frac{dw}{dx} - \frac{dw}{dx}\bigg|_0\right) = P\left(Lx - \frac{1}{2}x^2\right) \qquad (84)$$

Now based on the fact that if the value of $\dfrac{dw}{dx}\bigg|_0$ is known or not, we consider partially or fully clamped cases in the following subsections.

### 5.1. Partially clamped cantilever beam

In this case the slope $\dfrac{dw}{dx}\bigg|_0$ is not yet known. Therefore, equation (84) can be written as

$$\frac{dw}{dx} - \frac{dw}{dx}\bigg|_0 = \frac{P}{2k_\mu GAl^2}\left(Lx - \frac{1}{2}x^2\right) - \frac{EI + 2k_\mu GAl^2}{2k_\mu GAl^2}\beta \qquad (85)$$

By some manipulation, we obtain the relation

$$\frac{dw}{dx} - \frac{k_\mu}{k_s}l^2\frac{d^3w}{dx^3} = \frac{dw}{dx}\bigg|_0 + \frac{P}{2k_\mu GAl^2}\left(Lx - \frac{1}{2}x^2 + \frac{k_\mu}{k_s}l^2\right) - \frac{\left(EI + 2k_\mu GAl^2\right)}{2k_\mu GAl^2}\left(\beta - \frac{k_\mu}{k_s}l^2\frac{d^2\beta}{dx^2}\right) \qquad (86)$$

By using this relation in (80), we obtain the second order linear differential equation for $\beta$ as

$$\left(EI + 4k_\mu GAl^2\right)\beta - EI\frac{k_\mu}{k_s}l^2\frac{d^2\beta}{dx^2} = 2k_\mu GAl^2\frac{dw}{dx}\bigg|_0 + P\left(Lx - \frac{1}{2}x^2\right) - P\frac{k_\mu}{k_s}l^2 \qquad (87)$$

By defining the new length scale parameter

$$\ell = l\sqrt{\frac{k_\mu}{k_s}\frac{EI}{EI + 4k_\mu GAl^2}} \qquad (88)$$

the solution to this boundary value problem for $\beta$ is given by

$$\beta = \frac{P}{EI + 4k_\mu GAl^2}\left(Lx - \frac{1}{2}x^2\right) + a_0\left[1 - \frac{\cosh\left(\frac{L-x}{\ell}\right)}{\cosh\left(\frac{L}{\ell}\right)}\right] \tag{89}$$

where we have used the boundary condition $\beta = 0$ at $x = 0$. By using the boundary condition $M_\sigma = EI\frac{d\beta}{dx} = 0$ at $x = L$, we obtain for the coefficient $a_0$

$$a_0 = -2P\frac{k_\mu}{k_s}l^2\frac{EI + 2k_\mu GAl^2}{\left(EI + 4k_\mu GAl^2\right)^2} + \frac{2k_\mu GAl^2}{EI + 4k_\mu GAl^2}\frac{dw}{dx}\bigg|_0 \tag{90}$$

The value of $\frac{dw}{dx}\bigg|_0$ is obtained by enforcing the boundary condition $M_\mu = 2k_\mu Gl^2\left(\frac{d^2w}{dx^2} + \frac{d\beta}{dx}\right) = 0$

at $x = 0$. By using the expressions (85) and (89), we obtain

$$\frac{d^2w}{dx^2} + \frac{d\beta}{dx} = \frac{P}{2k_\mu GAl^2}(L - x) - \frac{EI}{2k_\mu GAl^2}\frac{d\beta}{dx} \tag{91}$$

which shows that

$$\frac{d\beta}{dx} = \frac{PL}{EI} \qquad \text{at} \qquad x = 0 \tag{92}$$

By using this boundary value in (89), we obtain the explicit relation

$$a_0 = PL\frac{4k_\mu GAl^2}{EI\left(EI + 4k_\mu GAl^2\right)}\frac{\ell}{\tanh\left(\frac{L}{\ell}\right)} \tag{93}$$

As a result, the relation (90) gives

$$\frac{dw}{dx}\bigg|_0 = \frac{2PL}{EI}\frac{\ell}{\tanh\left(\frac{L}{\ell}\right)} + \frac{P}{k_s GA}\frac{EI + 2k_\mu GAl^2}{EI + 4k_\mu GAl^2} \tag{94}$$

Consequently, for $\beta$ and $\theta = \frac{dw}{dx}$, we obtain

$$\beta = \frac{P}{EI+4k_\mu GAl^2}\left(Lx-\frac{1}{2}x^2\right)+PL\frac{4k_\mu GAl^2}{EI(EI+4k_\mu GAl^2)}\frac{\ell}{\tanh\left(\frac{L}{\ell}\right)}\left[1-\frac{\cosh\left(\frac{L-x}{\ell}\right)}{\cosh\left(\frac{L}{\ell}\right)}\right] \quad (95)$$

$$\theta = \frac{dw}{dx} = \frac{P}{k_s GA}\frac{EI+2k_\mu GAl^2}{EI+4k_\mu GAl^2}+\frac{2PL}{EI}\frac{\ell}{\tanh\left(\frac{L}{\ell}\right)}$$
$$+\frac{P}{EI+4k_\mu GAl^2}\left(Lx-\frac{1}{2}x^2\right)-2\frac{PL}{EI}\frac{EI+2k_\mu GAl^2}{EI+4k_\mu GAl^2}\frac{\ell}{\tanh\left(\frac{L}{\ell}\right)}\left[1-\frac{\cosh\left(\frac{L-x}{\ell}\right)}{\cosh\left(\frac{L}{\ell}\right)}\right] \quad (96)$$

By integrating the relation (96) and noticing $w=0$ at $x=0$, we obtain the expression for the transverse deflection $w$ as

$$w = \left[\frac{P}{k_s GA}\frac{EI+2k_\mu GAl^2}{EI+4k_\mu GAl^2}+\frac{2PL}{EI}\frac{\ell}{\tanh\left(\frac{L}{\ell}\right)}\right]x+\frac{P}{EI+4k_\mu GAl^2}\left(\frac{1}{2}Lx^2-\frac{1}{6}x^3\right)$$
$$-\frac{2PL}{EI}\frac{EI+2k_\mu GAl^2}{EI+4k_\mu GAl^2}\frac{\ell}{\tanh\left(\frac{L}{\ell}\right)}\left[x+\ell\frac{\sinh\left(\frac{L-x}{\ell}\right)}{\cosh\left(\frac{L}{\ell}\right)}-\ell\tanh\left(\frac{L}{\ell}\right)\right] \quad (97)$$

The general deformation of this partially clamped cantilever beam is shown in Fig. 5.

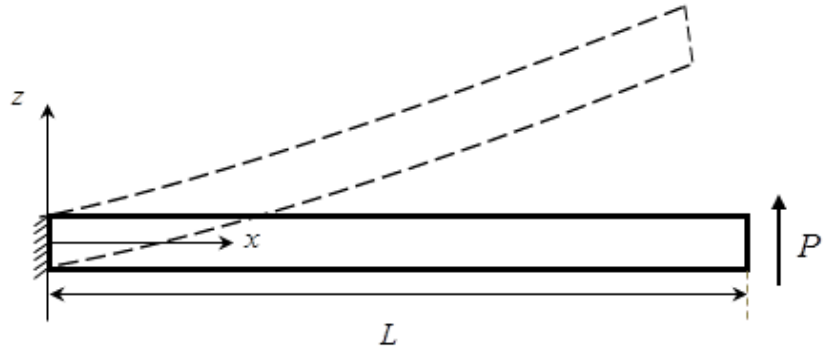

**Fig. 5.** Transverse deformation for partially clamped cantilever beam.

For the corresponding engineering shear strain $\gamma$, rotation $\phi$ and engineering curvature $\mathcal{K}$, we obtain

$$\gamma = \frac{dw}{dx} - \beta = \frac{P}{k_s GA} \frac{EI + 2k_\mu GAl^2}{EI + 4k_\mu GAl^2} + 2\frac{PL}{EI} \ell \frac{\cosh\left(\frac{L-x}{\ell}\right)}{\sinh\left(\frac{L}{\ell}\right)} \tag{98}$$

$$\phi = \frac{1}{2}\left(\frac{dw}{dx} + \beta\right) = \frac{PL}{EI} \frac{\ell}{\tanh\left(\frac{L}{\ell}\right)} + \frac{P}{2k_s GA} \frac{EI + 2k_\mu GAl^2}{EI + 4k_\mu GAl^2}$$

$$+ \frac{P}{EI + 4k_\mu GAl^2}\left(Lx - \frac{1}{2}x^2\right) - \frac{PL}{EI + 4k_\mu GAl^2} \frac{\ell}{\tanh\left(\frac{L}{\ell}\right)}\left[1 - \frac{\cosh\left(\frac{L-x}{\ell}\right)}{\cosh\left(\frac{L}{\ell}\right)}\right] \tag{99}$$

$$\mathcal{K} = \frac{d\phi}{dx} = \frac{P}{EI + 4k_\mu GAl^2}(L-x) - \frac{PL}{EI + 4k_\mu GAl^2} \frac{\sinh\left(\frac{L-x}{\ell}\right)}{\sinh\left(\frac{L}{\ell}\right)} \tag{100}$$

As a result, the force- and couple-stress bending moments become

$$M_\sigma = EI \frac{d\beta}{dx}$$

$$= \frac{EI}{EI + 4k_\mu GAl^2} P(L-x) + \frac{4k_\mu GAl^2}{EI + 4k_\mu GAl^2} PL \frac{\sinh\left(\frac{L-x}{\ell}\right)}{\sinh\left(\frac{L}{\ell}\right)} \tag{101}$$

$$M_\mu = 4k_\mu GAl^2 \frac{d\phi}{dx}$$

$$= \frac{4k_\mu GAl^2}{EI + 4k_\mu GAl^2} P(L-x) - \frac{4k_\mu GAl^2}{EI + 4k_\mu GAl^2} PL \frac{\sinh\left(\frac{L-x}{\ell}\right)}{\sinh\left(\frac{L}{\ell}\right)} \tag{102}$$

For the shear and couple-stress induced transverse forces, we obtain

$$Q_\sigma = k_s GA\gamma$$

$$= P\frac{EI + 2k_\mu GAl^2}{EI + 4k_\mu GAl^2} + 2P\frac{k_s GAL}{EI}\ell\,\frac{\cosh\left(\frac{L-x}{\ell}\right)}{\sinh\left(\frac{L}{\ell}\right)} \tag{103}$$

$$Q_\mu = -\frac{1}{2}\frac{dM_\mu}{dx}$$

$$= P\frac{2k_\mu GAl^2}{EI + 4k_\mu GAl^2} - 2P\frac{k_s GAL}{EI}\ell\,\frac{\cosh\left(\frac{L-x}{\ell}\right)}{\sinh\left(\frac{L}{\ell}\right)} \tag{104}$$

Interestingly, we notice that

$$Q_\sigma + Q_\mu = P = Q \tag{105}$$

$$M_\sigma + M_\mu = P(L-x) = M \tag{106}$$

For the tip deflection of the beam, we have

$$\delta = w(L) = \frac{PL^3}{3(EI + 4k_\mu GAl^2)}$$
$$+ \frac{PL}{k_s GA}\frac{EI + 2k_\mu GAl^2}{EI + 4k_\mu GAl^2} + 2PL\frac{EI + 2k_\mu GAl^2}{(EI + 4k_\mu GAl^2)^2}\frac{k_\mu}{k_s}l^2 + \frac{2PL^2}{EI}\frac{2k_\mu GAl^2}{EI + 4k_\mu GAl^2}\frac{\ell}{\tanh\left(\frac{L}{\ell}\right)} \tag{107}$$

Therefore, the spring constant or stiffness $K = \frac{P}{\delta}$ of the cantilever becomes

$$K = \frac{1}{\dfrac{L^3}{3(EI + 4k_\mu GAl^2)} + \dfrac{L}{k_s GA}\dfrac{EI + 2k_\mu GAl^2}{EI + 4k_\mu GAl^2} + 2L\dfrac{EI + 2k_\mu GAl^2}{(EI + 4k_\mu GAl^2)^2}\dfrac{k_\mu}{k_s}l^2 + \dfrac{2L^2}{EI}\dfrac{2k_\mu GAl^2}{EI + 4k_\mu GAl^2}\dfrac{\ell}{\tanh\left(\frac{L}{\ell}\right)}}$$

$$\tag{108}$$

We may define the effective flexural rigidity $E_{eff}I = \dfrac{KL^3}{3}$ based on the stiffness in classical Euler-Bernoulli beam theory as

$$E_{eff}I = \frac{1}{3}KL^3$$

$$= \frac{1}{\dfrac{1}{EI+4k_\mu GAl^2} + \dfrac{3}{k_s GAL^2}\dfrac{EI+2k_\mu GAl^2}{EI+4k_\mu GAl^2} + \dfrac{6}{L^2}\dfrac{EI+2k_\mu GAl^2}{\left(EI+4k_\mu GAl^2\right)^2}\dfrac{k_\mu}{k_s}l^2 + \dfrac{6}{EI}\dfrac{2k_\mu GAl^2}{EI+4k_\mu GAl^2}\dfrac{\dfrac{\ell}{L}}{\tanh\left(\dfrac{L}{\ell}\right)}}$$

(109)

Therefore, the non-dimensional stiffness or flexural rigidity $R$ is

$$R = \frac{E_{eff}I}{EI} = \frac{KL^3}{3EI}$$

$$= \frac{1}{\dfrac{EI}{EI+4k_\mu GAl^2} + \dfrac{3EI}{k_s GAL^2}\dfrac{EI+2k_\mu GAl^2}{EI+4k_\mu GAl^2} + \dfrac{6EI}{L^2}\dfrac{EI+2k_\mu GAl^2}{\left(EI+4k_\mu GAl^2\right)^2}\dfrac{k_\mu}{k_s}l^2 + \dfrac{12k_\mu GAl^2}{EI+4k_\mu GAl^2}\dfrac{\dfrac{\ell}{L}}{\tanh\left(\dfrac{L}{\ell}\right)}}$$

(110)

We notice that when the length of beam becomes larger and larger, the effective flexural rigidity approaches its value in couple stress Euler-Bernoulli beam theory, where

$$L \to \infty \qquad \left(E_{eff}I\right)_{L=\infty} = EI + 4k_\mu GAl^2 \qquad (111)$$

This means that for longer beams the transverse deformation of the beam is negligible. Now we consider the limiting cases based on the length scale parameter $l$ as follows.

### 5.1.1. Classical theory $l = 0$

When the couple-stress effects are neglected, that is $l = 0$, we notice that $\ell = 0$. As a result, the solution reduces to the classical Timoshenko solution, where

$$w = \frac{P}{EI}\left(\frac{1}{2}Lx^2 - \frac{1}{6}x^3\right) + \frac{P}{k_s GA}x \qquad (112)$$

$$\theta = \frac{dw}{dx} = \frac{P}{EI}\left(Lx - \frac{1}{2}x^2\right) + \frac{P}{k_s GA} \qquad (113)$$

$$\beta = \frac{P}{EI + 4k_\mu GAl^2}\left(Lx - \frac{1}{2}x^2\right) \tag{114}$$

$$\gamma = \frac{P}{k_s GA} \tag{115}$$

$$\phi = \frac{1}{2}\left(\frac{dw}{dx} + \beta\right) = \frac{P}{EI}\left(Lx - \frac{1}{2}x^2\right) + \frac{P}{2k_s GA} \tag{116}$$

Interestingly, we notice

$$Q_\sigma = P = Q \tag{117}$$

$$M_\sigma = P(L-x) = M \tag{118}$$

For the stiffness and the effective flexural rigidity, we obtain

$$K_0 = \frac{1}{\dfrac{L^3}{3EI} + \dfrac{L}{k_s GA}} \tag{119}$$

$$\left(E_{eff}I\right)_0 = \frac{K_0 L^3}{3} = \frac{EI}{1 + \dfrac{3EI}{k_s GAL^2}} \tag{120}$$

We should notice that the effective flexural rigidity $\left(E_{eff}I\right)_0$ in classical Timoshenko beam theory is apparently size-dependent and less than its corresponding value $EI$ in classical Euler-Bernoulli beam theory. This classical size-dependency is the result of the peculiar definition of $\left(E_{eff}I\right)_0$ that includes the effect of shear deformation in classical beam theory. For this case, the normalized effective flexural rigidity is

$$R_0 = \frac{\left(E_{eff}I\right)_0}{EI} = \frac{1}{1 + \dfrac{3EI}{k_s GAL^2}} \tag{121}$$

which is less than one, that is $R_0 < 1$.

*5.1.2. Inflexurable couple stress theory* $l \to \infty$

When all dimensions of the beam are smaller than the length scale parameter $l$, the couple-stress effect becomes dominant. Interestingly, we can consider the limiting case $l \to \infty$, where the engineering curvature $\mathbb{K} \to 0$ and $\ell \to \ell_\infty$, with

$$\ell_\infty = \frac{1}{2}\sqrt{\frac{EI}{k_s GA}} \tag{122}$$

This interesting case may be called inflexurable, where there is no continuum mechanical curvature. Therefore, the solution reduces to

$$w = \left[\frac{P}{2k_s GA} + \frac{PL}{EI}\frac{\ell_\infty}{\tanh\left(\frac{L}{\ell_\infty}\right)}\right]x - \frac{PL}{4k_s GA}\left[\frac{\sinh\left(\frac{L-x}{\ell_\infty}\right)}{\sinh\left(\frac{L}{\ell_\infty}\right)} - 1\right] \tag{123}$$

$$\theta = \frac{dw}{dx} = \frac{P}{2k_s GA} + \frac{PL}{EI}\frac{\ell_\infty}{\tanh\left(\frac{L}{\ell_\infty}\right)} + \frac{PL}{4k_s GA}\frac{1}{\ell_\infty}\frac{\cosh\left(\frac{L-x}{\ell_\infty}\right)}{\sinh\left(\frac{L}{\ell_\infty}\right)} \tag{124}$$

$$\beta = \frac{PL}{EI}\frac{\ell_\infty}{\tanh\left(\frac{L}{\ell_\infty}\right)}\left[1 - \frac{\cosh\left(\frac{L-x}{\ell_\infty}\right)}{\cosh\left(\frac{L}{\ell_\infty}\right)}\right] \tag{125}$$

$$\gamma = \frac{dw}{dx} - \beta = \frac{P}{2k_s GA} + 2\frac{PL}{EI}\frac{\ell_\infty}{\tanh\left(\frac{L}{\ell_\infty}\right)}\frac{\cosh\left(\frac{L-x}{\ell_\infty}\right)}{\cosh\left(\frac{L}{\ell_\infty}\right)} \tag{126}$$

$$\phi = \frac{1}{2}\left(\frac{dw}{dx} + \beta\right) = \frac{PL}{EI}\frac{\ell_\infty}{\tanh\left(\frac{L}{\ell_\infty}\right)} + \frac{P}{4k_s GA} \tag{127}$$

$$\mathbb{K} = \frac{d\phi}{dx} = 0 \tag{128}$$

We notice that although the slope $\theta = \dfrac{dw}{dx}$ and $\beta$ are not constant, the continuum mechanical rotation $\phi$ is constant and the engineering curvature $\mathcal{K}$ becomes zero. For this inflexurable case, the force- and couple-stress bending moments approach, respectively, to the forms

$$M_\sigma = EI\frac{d\beta}{dx} = PL\frac{\sinh\left(\dfrac{L-x}{\ell_\infty}\right)}{\sinh\left(\dfrac{L}{\ell_\infty}\right)} \tag{129}$$

$$M_\mu = P(L-x) - PL\frac{\sinh\left(\dfrac{L-x}{\ell_\infty}\right)}{\sinh\left(\dfrac{L}{\ell_\infty}\right)} \tag{130}$$

and the shear and couple-stress induced transverse forces become, respectively,

$$Q_\sigma = \frac{1}{2}P\left[1 + \frac{L}{\ell_\infty}\frac{\cosh\left(\dfrac{L-x}{\ell_\infty}\right)}{\sinh\left(\dfrac{L}{\ell_\infty}\right)}\right] \tag{131}$$

$$Q_\mu = \frac{1}{2}P\left[1 - \frac{L}{\ell_\infty}\frac{\cosh\left(\dfrac{L-x}{\ell_\infty}\right)}{\sinh\left(\dfrac{L}{\ell_\infty}\right)}\right] \tag{132}$$

The tip deflection of the beam approaches to

$$\delta_\infty = \frac{3PL}{4k_s GA} + \frac{PL^2}{EI}\frac{\ell_\infty}{\tanh\left(\dfrac{L}{\ell_\infty}\right)} \tag{133}$$

and the stiffness becomes

$$K_\infty = \frac{1}{\dfrac{3L}{4k_s GA} + \dfrac{L^2}{EI}\dfrac{\ell_\infty}{\tanh\left(\dfrac{L}{\ell_\infty}\right)}} \tag{134}$$

Therefore, for this case, the effective flexural rigidity and its normalized value are

$$(E_{eff}I)_\infty = \frac{K_\infty L^3}{3} = \frac{1}{\frac{9}{4k_s GAL^2} + \frac{3}{EIL}\frac{\ell_\infty}{\tanh\left(\frac{L}{\ell_\infty}\right)}} \tag{135}$$

$$R_\infty = \frac{(E_{eff}I)_\infty}{EI} = \frac{K_\infty L^3}{3EI} = \frac{1}{\frac{9EI}{4k_s GAL^2} + \frac{3}{L}\frac{\ell_\infty}{\tanh\left(\frac{L}{\ell_\infty}\right)}} \tag{136}$$

respectively.

When the beam is so long that $L \gg \ell_\infty$, we have $\tanh\left(\frac{L}{\ell_\infty}\right) \approx 1$. Therefore, the deformation can be approximately represented by

$$w = \left[\frac{P}{2k_s GA} + \frac{PL}{EI}\ell_\infty\right]x + \frac{PL}{4k_s GA}\left(1 - e^{-\frac{x}{\ell_\infty}}\right) \tag{137}$$

$$\theta = \frac{dw}{dx} = \frac{P}{2k_s GA} + \frac{PL}{EI}\ell_\infty + \frac{PL}{4k_s GA}\frac{1}{\ell_\infty}e^{-\frac{x}{\ell_\infty}} \tag{138}$$

$$\beta = \frac{PL}{EI}\ell_\infty\left(1 - e^{-\frac{x}{\ell_\infty}}\right) \tag{139}$$

$$Q_\sigma = \frac{P}{2}\left(1 + \frac{L}{\ell_\infty}e^{-\frac{x}{\ell_\infty}}\right) \tag{140}$$

$$Q_\mu = \frac{P}{2}\left(1 - \frac{L}{\ell_\infty}e^{-\frac{x}{\ell_\infty}}\right) \tag{141}$$

$$M_\sigma = P\ell_\infty e^{-\frac{x}{\ell_\infty}} \tag{142}$$

$$M_\mu = P(L-x) - P\ell_\infty e^{-\frac{x}{\ell_\infty}} \tag{143}$$

The relation (137) shows that except a small region near to the fixed end, the deflection is practically a straight line, where we have

$$w = \frac{PL}{4k_s GA} + \left(\frac{P}{2k_s GA} + \frac{PL}{EI}\ell_\infty\right)x \qquad x > 3\ell_\infty \qquad (144)$$

Therefore, we may approximate the relations (134)-(136) by the expressions

$$K_\infty = \frac{EI}{\ell_\infty L^2} = \frac{2}{L^2}\sqrt{k_s EIGA} \qquad (145)$$

$$\left(E_{eff} I\right)_\infty = \frac{EI}{3\ell_\infty}L = \frac{2}{3}L\sqrt{k_s EIGA} \qquad (146)$$

$$R_\infty = \frac{\left(E_{eff} I\right)_\infty}{EI} = \frac{1}{3\ell_\infty}L = \frac{2}{3}L\sqrt{\frac{k_s GA}{EI}} \qquad (147)$$

### 5.2. Fully clamped cantilever beam

In this case, there is no slope at $x=0$, that is $\left.\dfrac{dw}{dx}\right|_0 = 0$. As a result, the equation (85) gives

$$\frac{dw}{dx} = \frac{P}{2k_\mu GAl^2}\left(Lx - \frac{1}{2}x^2\right) - \frac{EI + 2k_\mu GAl^2}{2k_\mu GAl^2}\beta \qquad (148)$$

From this relation, we obtain

$$\frac{dw}{dx} - \frac{k_\mu}{k_s}l^2\frac{d^3w}{dx^3} = \frac{P}{2k_\mu GAl^2}\left(Lx - \frac{1}{2}x^2 + \frac{k_\mu}{k_s}l^2\right) - \frac{\left(EI + 2k_\mu GAl^2\right)}{2k_\mu GAl^2}\left(\beta - \frac{k_\mu}{k_s}l^2\frac{d^2\beta}{dx^2}\right) \qquad (149)$$

Therefore, by using this in (81), the second order linear differential equation for $\beta$ becomes

$$\left(EI + 4k_\mu GAl^2\right)\beta - EI\frac{k_\mu}{k_s}l^2\frac{d^2\beta}{dx^2} = P\left(Lx - \frac{1}{2}x^2\right) - P\frac{k_\mu}{k_s}l^2 \qquad (150)$$

By using the boundary conditions $\beta = 0$ at $x=0$, and $M_\sigma = EI\dfrac{d\beta}{dx} = 0$ at $x = L$, we obtain the solution as

$$\beta = \frac{P}{EI + 4k_\mu GAl^2}\left(Lx - \frac{1}{2}x^2\right) - 2P\frac{k_\mu}{k_s}l^2\frac{EI + 2k_\mu GAl^2}{\left(EI + 4k_\mu GAl^2\right)^2}\left[1 - \frac{\cosh\left(\frac{L-x}{\ell}\right)}{\cosh\left(\frac{L}{\ell}\right)}\right] \quad (151)$$

By using this expression $\beta$ in (148), we obtain the relation for the slope $\theta = \dfrac{dw}{dx}$ as

$$\theta = \frac{dw}{dx} = \frac{P}{EI + 4k_\mu GAl^2}\left(Lx - \frac{1}{2}x^2\right) + \frac{P}{k_s GA}\frac{\left(EI + 2k_\mu GAl^2\right)^2}{\left(EI + 4k_\mu GAl^2\right)^2}\left[1 - \frac{\cosh\left(\frac{L-x}{\ell}\right)}{\cosh\left(\frac{L}{\ell}\right)}\right] \quad (152)$$

By integrating this relation and noticing $w = 0$ at $x = 0$, we obtain the transverse deflection $w$ as

$$w = \frac{P}{EI + 4k_\mu GAl^2}\left(\frac{1}{2}Lx^2 - \frac{1}{6}x^3\right) + \frac{P}{k_s GA}\frac{\left(EI + 2k_\mu GAl^2\right)^2}{\left(EI + 4k_\mu GAl^2\right)^2}\left[x - \ell\tanh\left(\frac{L}{\ell}\right) + \ell\frac{\sinh\left(\frac{L-x}{\ell}\right)}{\cosh\left(\frac{L}{\ell}\right)}\right] \quad (153)$$

Figure 6 shows the general deformation of this fully clamped cantilever beam.

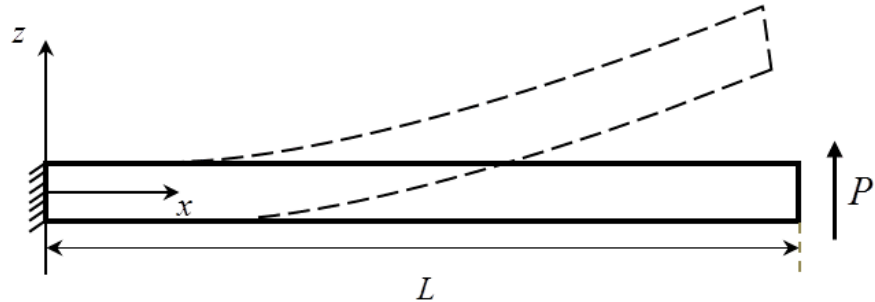

**Fig. 6.** Transverse deformation for fully clamped cantilever beam.

For this deformation, the corresponding engineering shear strain $\gamma$, rotation $\phi$ and engineering curvature $\mathbb{K}$ are

$$\gamma = \frac{dw}{dx} - \beta = \frac{P}{k_s GA} \frac{EI + 2k_\mu GAl^2}{EI + 4k_\mu GAl^2}\left[1 - \frac{\cosh\left(\frac{L-x}{\ell}\right)}{\cosh\left(\frac{L}{\ell}\right)}\right] \tag{154}$$

$$\phi = \frac{1}{2}\left(\frac{dw}{dx} + \beta\right) = \frac{P}{EI + 4k_\mu GAl^2}\left(Lx - \frac{1}{2}x^2\right) + \frac{P}{2}\frac{EI}{k_s GA}\frac{EI + 2k_\mu GAl^2}{\left(EI + 4k_\mu GAl^2\right)^2}\left[1 - \frac{\cosh\left(\frac{L-x}{\ell}\right)}{\cosh\left(\frac{L}{\ell}\right)}\right] \tag{155}$$

$$\kappa = \frac{d\phi}{dx} = \frac{P}{EI + 4k_\mu GAl^2}\left(L - \frac{1}{2}x\right) + \frac{P}{2}\frac{EI}{k_s GA}\frac{EI + 2k_\mu GAl^2}{\left(EI + 4k_\mu GAl^2\right)^2}\frac{1}{\ell}\frac{\sinh\left(\frac{L-x}{\ell}\right)}{\cosh\left(\frac{L}{\ell}\right)} \tag{156}$$

For the force- and couple-stress bending moment, we have, respectively,

$$M_\sigma = EI\frac{d\beta}{dx}$$
$$= \frac{EI}{EI + 4k_\mu GAl^2}P(L-x) - 2P\frac{k_\mu}{k_s}\frac{l^2}{\ell}\frac{EI(EI + 2k_\mu GAl^2)}{(EI + 4k_\mu GAl^2)^2}\frac{\sinh\left(\frac{L-x}{\ell}\right)}{\cosh\left(\frac{L}{\ell}\right)} \tag{157}$$

$$M_\mu = 4k_\mu GAl^2\kappa$$
$$= \frac{4k_\mu GAl^2}{EI + 4k_\mu GAl^2}P(L-x) + 2P\frac{k_\mu}{k_s}\frac{l^2}{\ell}\frac{EI(EI + 2k_\mu GAl^2)}{(EI + 4k_\mu GAl^2)^2}\frac{\sinh\left(\frac{L-x}{\ell}\right)}{\cosh\left(\frac{L}{\ell}\right)} \tag{158}$$

For the shear and couple-stress induced transverse forces, we obtain

$$Q_\sigma = k_s GA\gamma = P\frac{EI + 2k_\mu GAl^2}{EI + 4k_\mu GAl^2}\left[1 - \frac{\cosh\left(\frac{L-x}{\ell}\right)}{\cosh\left(\frac{L}{\ell}\right)}\right] \tag{159}$$

$$Q_\mu = -\frac{1}{2}\frac{dM_\mu}{dx}$$

$$= P\frac{2k_\mu GAl^2}{EI + 4k_\mu GAl^2} - P\frac{EI + 2k_\mu GAl^2}{EI + 4k_\mu GAl^2}\frac{\cosh\left(\frac{L-x}{\ell}\right)}{\cosh\left(\frac{L}{\ell}\right)} \qquad (160)$$

where

$$Q_\sigma + Q_\mu = P = Q \qquad (161)$$

$$M_\sigma + M_\mu = P(L-x) = M \qquad (162)$$

For the tip deflection of the beam, we obtain

$$\delta = w(L) = \frac{PL^3}{3}\frac{1}{EI + 4k_\mu GAl^2} + \frac{P}{k_s GA}\frac{\left(EI + 2k_\mu GAl^2\right)^2}{\left(EI + 4k_\mu GAl^2\right)^2}\left[L - \ell\tanh\left(\frac{L}{\ell}\right)\right] \qquad (163)$$

As a result, for the spring constant or stiffness of the cantilever in this case, we obtain

$$K = \frac{1}{\dfrac{L^3}{3\left(EI + 4k_\mu GAl^2\right)} + \dfrac{L}{k_s GA}\dfrac{\left(EI + 2k_\mu GAl^2\right)^2}{\left(EI + 4k_\mu GAl^2\right)^2}\left[1 - \dfrac{\ell}{L}\tanh\left(\dfrac{L}{\ell}\right)\right]} \qquad (164)$$

For the effective flexural rigidity based on the stiffness in classical Euler-Bernoulli beam theory, we have

$$E_{eff}I = \frac{KL^3}{3}$$

$$= \frac{1}{\dfrac{1}{EI + 4k_\mu GAl^2} + \dfrac{1}{k_s GA}\dfrac{3}{L^2}\dfrac{\left(EI + 2k_\mu GAl^2\right)^2}{\left(EI + 4k_\mu GAl^2\right)^2}\left[1 - \dfrac{\ell}{L}\tanh\left(\dfrac{L}{\ell}\right)\right]} \qquad (165)$$

Therefore, the non-dimensional stiffness or flexural rigidity becomes

$$R = \frac{E_{eff}I}{EI} = \frac{KL^3}{3EI}$$

$$= \frac{1}{\dfrac{EI}{EI + 4k_\mu GAl^2} + \dfrac{EI}{k_s GA}\dfrac{3}{L^2}\dfrac{\left(EI + 2k_\mu GAl^2\right)^2}{\left(EI + 4k_\mu GAl^2\right)^2}\left[1 - \dfrac{\ell}{L}\tanh\left(\dfrac{L}{\ell}\right)\right]} \qquad (166)$$

Figure 7 shows the size-dependency of the non-dimensional flexural rigidity $R$ on the non-dimensional length $L/\sqrt{A}$. We notice that when the length of beam becomes larger and larger, the effective flexural rigidity approach to its corresponding value in couple stress Euler-Bernoulli beam theory, where

$$L \to \infty \qquad \left(E_{eff}I\right)_{L=\infty} = EI + 4k_\mu GAl^2 \qquad (167)$$

and therefore

$$R_{L=\infty} = \frac{\left(E_{eff}I\right)_{L=\infty}}{EI} = 1 + \frac{4k_\mu GAl^2}{EI} \qquad (168)$$

Similar to the partially clamped cantilever beam, for longer beams the transverse deformation of the beam is negligible. However, we notice that in size-dependent theory, the criteria for a long beam is much longer than the criteria in classical theory.

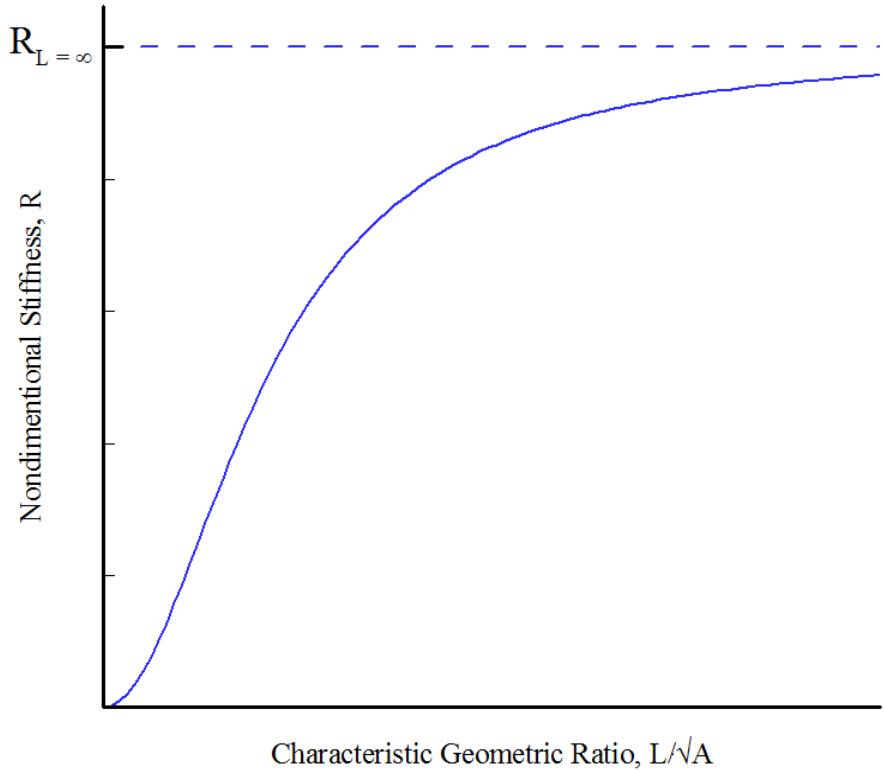

**Fig. 7.** Non-dimensional size-dependency of stiffness for fully clamped cantilever beam on non-dimensional length.

For this fully clamped case, we also consider the limiting cases based on the length scale parameter $l$ as follows.

*5.2.1. Classical theory $l = 0$*

When the couple-stress effects are neglected, that is $l = 0$, the solution reduces to the classical Timoshenko solution as discussed before in Section 5.1.1.

*5.2.2. Inflexurable couple stress theory $l \to \infty$*

For this case, there is no continuum mean curvature, that is $\mathbb{K} \to 0$. Consequently, the solution reduces to a pure transverse deformation, where

$$w = \frac{P}{4k_s GA}\left[x - \ell_\infty \tanh\left(\frac{L}{\ell_\infty}\right) + \ell_\infty \frac{\sinh\left(\frac{L-x}{\ell_\infty}\right)}{\cosh\left(\frac{L}{\ell_\infty}\right)}\right] \qquad (169)$$

$$\theta = \frac{dw}{dx} = \frac{P}{4k_s GA}\left[1 - \frac{\cosh\left(\frac{L-x}{\ell_\infty}\right)}{\cosh\left(\frac{L}{\ell_\infty}\right)}\right] \qquad (170)$$

$$\beta = -\frac{dw}{dx} = -\frac{P}{4k_s GA}\left[1 - \frac{\cosh\left(\frac{L-x}{\ell_\infty}\right)}{\cosh\left(\frac{L}{\ell_\infty}\right)}\right] \qquad (171)$$

$$\phi = \frac{1}{2}\left(\frac{dw}{dx} + \beta\right) = 0 \qquad (172)$$

$$\mathbb{K} = \frac{d\phi}{dx} = 0 \qquad (173)$$

$$\gamma = 2\frac{dw}{dx} = \frac{P}{2k_s GA}\left[1 - \frac{\cosh\left(\frac{L-x}{\ell_\infty}\right)}{\cosh\left(\frac{L}{\ell_\infty}\right)}\right] \tag{174}$$

We notice that although the slope $\theta = \dfrac{dw}{dx}$ and $\beta$ exist, they are equal and opposite of each other. As a result, the continuum mechanical rotation $\phi$ and engineering curvature $\kappa$ become zero. This means the inflexurable micro-beam is infinitely rigid in bending and can deform only by vertical motion of the cross-section, as illustrated in Fig. 8. However, we notice that this transverse deformation is not a classical shear deformation.

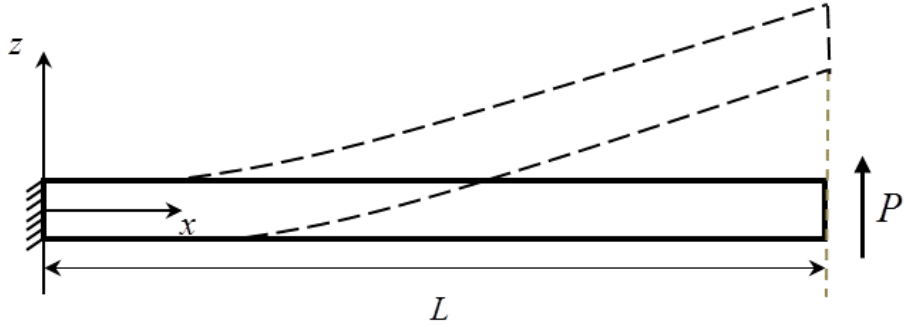

**Fig. 8.** Transverse deformation for fully clamped inflexurable beam, $l \to \infty$ or $\ell = \ell_\infty$.

For this case, the force- and couple-stress bending moments approach, respectively, to

$$M_\sigma = -P\ell_\infty \frac{\sinh\left(\frac{L-x}{\ell_\infty}\right)}{\cosh\left(\frac{L}{\ell_\infty}\right)} \tag{175}$$

$$M_\mu = P(L-x) + P\ell_\infty \frac{\sinh\left(\frac{L-x}{\ell_\infty}\right)}{\cosh\left(\frac{L}{\ell_\infty}\right)} \tag{176}$$

and the shear and couple-stress induced transverse forces become, respectively,

$$Q_\sigma = k_s GA\gamma = \frac{P}{2}\left[1 - \frac{\cosh\left(\frac{L-x}{\ell_\infty}\right)}{\cosh\left(\frac{L}{\ell_\infty}\right)}\right] \qquad (177)$$

$$Q_\mu = -\frac{1}{2}\frac{dM_\mu}{dx} = \frac{P}{2}\left[1 + \frac{\cosh\left(\frac{L-x}{\ell_\infty}\right)}{\cosh\left(\frac{L}{\ell_\infty}\right)}\right] \qquad (178)$$

Accordingly, the tip-deflection of the beam approaches to

$$\delta_\infty = \frac{P}{4k_s GA}\left[L - \ell_\infty \tanh\left(\frac{L}{\ell_\infty}\right)\right] \qquad (179)$$

and the stiffness becomes

$$K_\infty = \frac{4k_s GA}{L}\frac{1}{1-\frac{\ell_\infty}{L}\tanh\left(\frac{L}{\ell_\infty}\right)} \qquad (180)$$

Therefore, for this case, the effective flexural rigidity and its normalized value approach to

$$\left(E_{eff}I\right)_\infty = \frac{K_\infty L^3}{3} = \frac{4}{3}k_s GAL^2 \frac{1}{1-\frac{\ell_\infty}{L}\tanh\left(\frac{L}{\ell_\infty}\right)} \qquad (181)$$

$$R_\infty = \frac{\left(E_{eff}I\right)_\infty}{EI} = \frac{K_\infty L^3}{3EI} = \frac{4k_s GA}{3EI}L^2\frac{1}{1-\frac{\ell_\infty}{L}\tanh\left(\frac{L}{\ell_\infty}\right)} \qquad (182)$$

When the beam is so long that $L \gg \ell_\infty$, the deformation can be approximately represented by

$$w = \frac{P}{4k_s GA}\left[x - \ell_\infty\left(1 - e^{-\frac{x}{\ell_\infty}}\right)\right] \qquad (183)$$

$$\theta = \frac{dw}{dx} = \frac{P}{4k_s GA}\left(1 - e^{-\frac{x}{\ell_\infty}}\right) \qquad (184)$$

$$\beta = -\frac{dw}{dx} = -\frac{P}{4k_s GA}\left(1 - e^{-\frac{x}{\ell_\infty}}\right) \tag{185}$$

$$Q_\sigma = \frac{P}{2}\left(1 - e^{-\frac{x}{\ell_\infty}}\right) \tag{186}$$

$$Q_\mu = \frac{P}{2}\left(1 + e^{-\frac{x}{\ell_\infty}}\right) \tag{187}$$

$$M_\sigma = -P\ell_\infty e^{-\frac{x}{\ell_\infty}} \tag{188}$$

$$M_\mu = P(L - x) + P\ell_\infty e^{-\frac{x}{\ell_\infty}} \tag{189}$$

This relation shows that except a small region near to the fixed end, the deflection is practically a straight line, where we have

$$w = \frac{P}{4k_s GA}(x - \ell_\infty) \qquad x > 3\ell_\infty \tag{190}$$

and

$$M_\mu = P(L - x) \qquad x > 3\ell_\infty \tag{191}$$

This result indicates that the bending moment load is carried almost entirely by couple-stresses, except in the small region near to the fixed end. For this case $\dfrac{1}{1 - \dfrac{\ell_\infty}{L}\tanh\left(\dfrac{L}{\ell_\infty}\right)} \approx 1$ in the expressions (180)-(182) and the following approximation can be used:

$$K_\infty = \frac{4k_s GA}{L} \tag{192}$$

$$(E_{\text{eff}} I)_\infty = \frac{K_\infty L^3}{3} = \frac{4}{3} k_s GA L^2 \tag{193}$$

$$R_\infty = \frac{(E_{\text{eff}} I)_\infty}{EI} = \frac{4k_s GA}{3EI} L^2 \tag{194}$$

## 6. Deformation of a wide rectangular cantilever beam

Now we examine the analytical results for bending of a wide cantilever beam of rectangular cross-section with width $b$ and height $h$, shown in Fig. 9. As explained in Section 3, we only need to replace the Young's modulus $E$ by $\dfrac{E}{1-\nu^2}$ in the flexural rigidity terms $EI$ and $E_{eff}I$. For this beam, the area and second moment of area are $A = bh$ and $I = \dfrac{1}{12}bh^3$, respectively. The shear coefficient factor can be taken as $k_s = \dfrac{5}{6}$. Since the distribution of the couple-stress $\mu_{xy}$ is almost uniform, we can take $k_\mu = 1$.

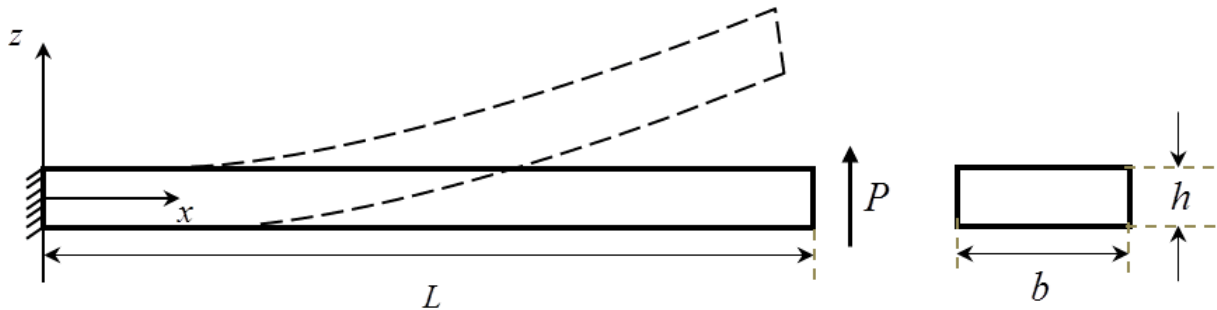

**Fig. 9.** Bending of Cantilever for a wide rectangular beam.

Here we investigate the size-dependency of non-dimensional stiffness or flexural rigidity $R = \dfrac{E_{eff}I}{EI} = \dfrac{KL^3}{3EI}$ on $h/l$ for partially and fully clamped boundary conditions. This problem has been also considered by Darrall et al. [22] in the framework of a two-dimensional finite element formulation method for a completely compressible material, where $\nu = 0$. Figures 10 and 11 show the size-dependency of non-dimensional stiffness $R$ on $h/l$ in the framework of size-dependent couple-stress Timoshenko beam theory for two different lengths, $L = 20h$ and $L = 40h$, for partially and fully clamped boundary conditions, respectively. We notice that these results for the completely compressible material, $\nu = 0$, are in full agreement with those in [22]. This clearly demonstrates the accuracy and consistency of the new size-dependent Timoshenko beam model with the underlying continuum theory.

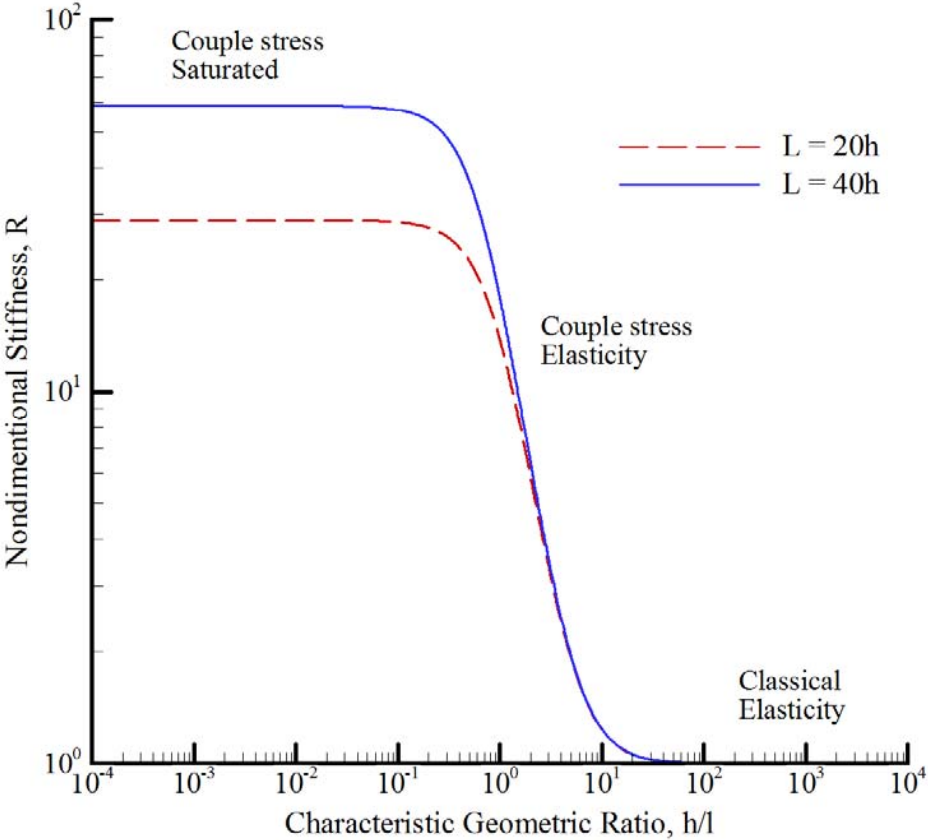

**Fig. 10.** Non-dimensional size-dependency of cantilever stiffness for partially clamped cantilever beam (completely compressible material, $v=0$).

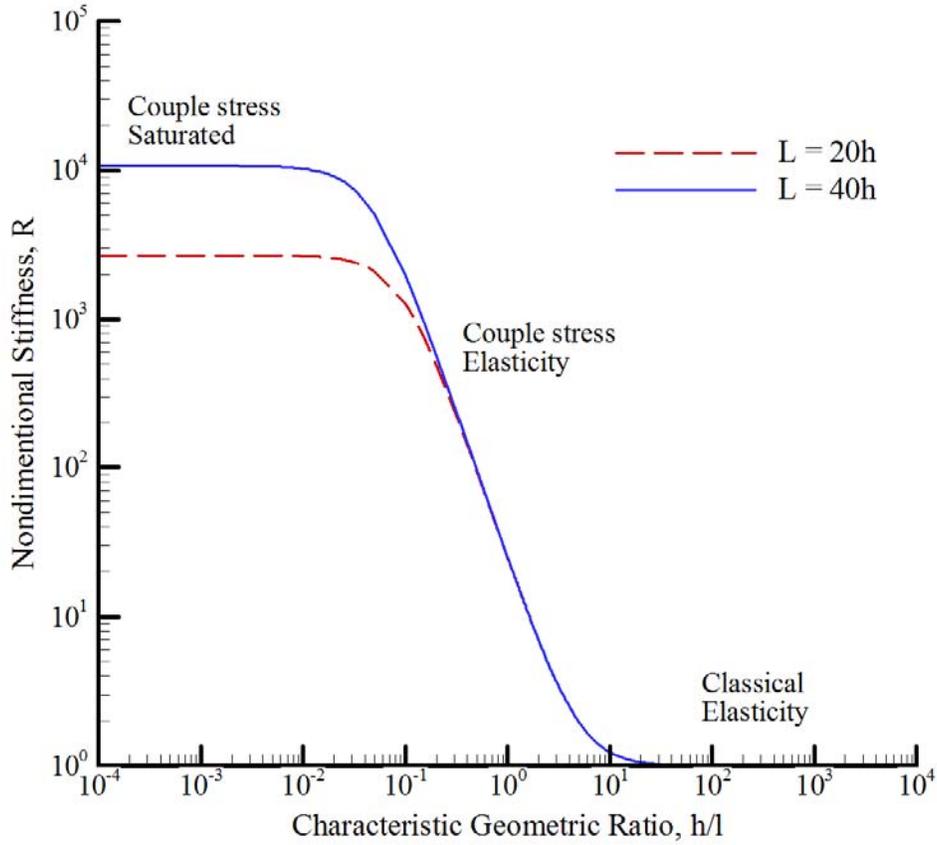

**Fig. 11.** Non-dimensional size-dependency of cantilever stiffness for fully clamped cantilever beam (completely compressible material, $\nu = 0$).

As noticed in [22], we can clearly see three well-defined domains associated with characteristic problem geometry as follows:

1. For large scale problems, where the characteristic geometry $h$ is much greater than $l$, that is, $h \gg l$, we have the classical elasticity region with stiffness independent of length scale. In this domain, couple-stress effects are negligible, mainly due to the small magnitude of curvature deformation at this scale. Notice that in this region the stiffness $K$ and non-dimensional stiffness $R$ can be approximated by their classical values $K_0$ and $R_0$, given by (119) and (121), respectively. Therefore, the stiffness and non-dimensional stiffness in this domain become

$$K_0 = \frac{3EI}{L^3} \frac{1}{1 + \frac{Eh^2}{4k_s G} \frac{h^2}{L^2}} \;, \qquad R_0 = \frac{1}{1 + \frac{Eh^2}{4k_s G} \frac{h^2}{L^2}} \qquad (195)$$

We notice that for this case, when the beam is very long, i.e., $L \gg h$, these relations for $K_0$ and $R_0$ approach the corresponding values in classical Euler-Bernoulli beam theory

$$K_0 = \frac{3EI}{L^3} \;, \qquad R_0 = 1 \qquad (196)$$

2. When the characteristic geometry for this problem is on the order of $l$, we enter the transitional couple-stress domain. For this cantilever problem, it is clear from Figs. 10 and 11 that couple-stress effects become significant for characteristic geometry with $h/l = 10$. In this couple-stress domain, there is an increase in flexural stiffness, which has a significant effect on the overall effective stiffness $K$ of the body.

3. Finally, for very small values of $h/l$, i.e., $h \ll l$, we have a domain that is couple-stress "saturated" in both Figs. 10 and 11. This means the flexural stiffness due to couple-stress effects has increased to the level where mean curvature is suppressed. We notice that in this nearly inflexurable domain, the deformation of the cantilever beam can be approximated by the inflexurability condition, where

$$\ell_\infty = \frac{h}{4}\sqrt{\frac{E}{3k_s G}} = \frac{h}{2}\sqrt{\frac{1+\nu}{6k_s}} \qquad (197)$$

Curiously, when the beam is very long, i.e., $L \gg h$, we have $L \gg \ell_\infty$. As a result, the relations (145), (147), (192) and (194) give the stiffness and non-dimensional stiffness in the couple stress "saturated" domains as

$$K_\infty = \frac{Eb}{6}\sqrt{\frac{6k_s}{1+\nu}}\frac{h^2}{L^2}, \qquad R_\infty = \frac{1}{3}\sqrt{\frac{24k_s}{1+\nu}}\frac{L}{h} \qquad \text{partially clamped} \qquad (198)$$

$$K_\infty = 4k_s Gb\frac{h}{L}, \qquad R_\infty = \frac{8k_s}{1+\nu}\frac{L^2}{h^2} \qquad \text{fully clamped} \qquad (199)$$

For the completely compressible case $\nu = 0$ and $k_s = 5/6$, we obtain

$$K_\infty = \frac{Eb}{6}\sqrt{5}\frac{h^2}{L^2}, \qquad R_\infty = \frac{2}{3}\sqrt{5}\frac{L}{h} \qquad \text{partially clamped} \qquad (200)$$

$$K_\infty = \frac{10}{3}Gb\frac{h}{L}, \qquad R_\infty = \frac{20}{3}\frac{L^2}{h^2} \qquad \text{fully clamped} \qquad (201)$$

Therefore, for sufficiently small $h/l$ ratios, an increase in total stiffness is observed, where the stiffness scales with $\frac{1}{L^2}$ and $\frac{1}{L}$ for partially and fully clamped conditions, respectively. Interestingly, these results have been concluded by Darrall et al. [22] based on numerical experiments.

Figure 12 demonstrates the deformation of the cantilever beam for the inflexurability condition for the fully clamped case, where the continuum mechanical rotation $\phi$ and engineering curvature $\mathcal{K}$ are zero. However, we notice that this transverse deformation is not a classical shear deformation.

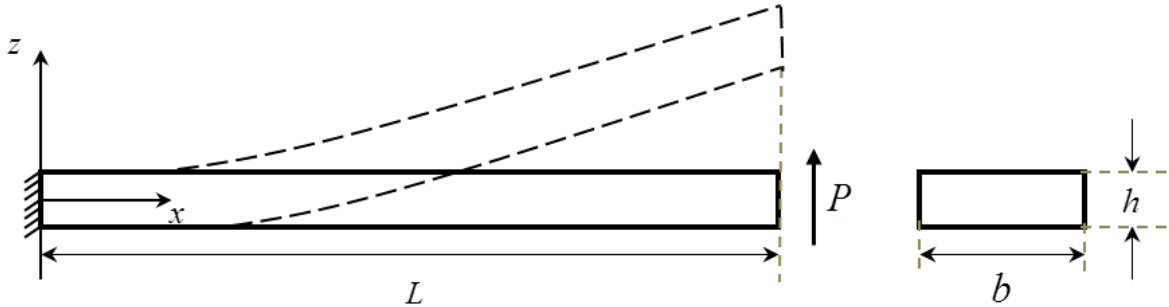

**Fig. 12.** Transverse deformation for fully clamped inflexurable material $l \to \infty$ $\ell = \ell_\infty$.

## 7. Conclusions

Based on consistent couple stress theory (C-CST), we have developed a size-dependent Timoshenko beam model. The corresponding governing equations and boundary conditions for this model have been obtained. We notice in this formulation the transverse displacement $w$, the slope $\theta = \dfrac{dw}{dx}$ and rotation $\beta$ are the degrees of freedom with energy conjugates $Q$, $M_\theta = \dfrac{1}{2} M_\mu$ and $M_\beta = M_\sigma + \dfrac{1}{2} M_\mu$, respectively. This clearly shows that the present couple stress Timoshenko beam theory combines the features of classical Timoshenko and Euler-Bernoulli beam theories.

Furthermore, this Timoshenko beam model has been employed to obtain analytical solutions for pure bending of a beam and for a cantilever beam with partially and fully clamped boundary conditions. For the cantilever beam, the limiting cases have been considered based on the length scale parameter $l$. The interesting inflexurable cases, corresponding to the limiting case $l \to \infty$, represent the condition for which there is no engineering curvature, $K = 0$. This approximates nearly inflexurable cases, where the material is almost rigid to curvature deformation. The beams with this condition result in saturated solutions, where the deformation is independent of the value of the length scale parameter $l$.

The inherent consistency and accuracy of the new size-dependent Timoshenko beam model has been demonstrated by comparing the analytical results to the numerical results from a two-dimensional finite element formulation of the corresponding couple stress continuum theory. This investigation shows that the analytical solutions based on the Timoshenko beam formulation correlate almost perfectly with the converged solutions obtained through mesh refinement of the finite element formulation.

Therefore, the self-consistency of C-CST makes it suitable for developing size-dependent structural models, such as beams, plates and shells to study both static and dynamic behavior, such as static deformation, buckling and vibration. However, the size-dependent modeling based on structural mechanics methods requires more approximation than the classical structural modeling, which does necessitate more careful attention to the underlying assumptions.